\documentclass[11pt]{article}
  \usepackage{amssymb}
   \usepackage{graphicx}
\renewcommand{\theequation}{\arabic{section}.\arabic{equation}}

\def\AAA{{\cal A}}  
 
\def\ba{\begin{eqnarray}}
\def\bam{\begin{array}}
\def\bA{\mbox{\bm$A\!\!$\ubm}~}
\def\bAAA{\mbox{\bm$\AAA\!$\ubm}~}

\def\bB{\mbox{\bm$B\!\!$\ubm}~}
\def\bBB{\mbox{\bm$\BB$\ubm}~}

\def\be{\begin{equation}}

\def\bE{\mbox{\bm$E\!$\ubm}~}

\def\bH{\bf{H}}
\def\bHH{\mbox{\bm$\HH$\ubm}~}
\def\bi{\bibitem}
\def\bJ{\mbox{\bm$J$\ubm}~}
\def\bl{\mbox{\bm$l\!\!$\ubm}~}
\def\bm{\boldmath}
\def\bn{\mbox{\bm$n\!\!$\ubm}~}
\def\bna{\mbox{\bm$\nabla\!\!$\ubm}~}

\def\bS{\mbox{\bm$S\!\!$\ubm}~}

\def\B {\overline}

\def\BB{{\cal B}}

\def\BC{\B C}

\def\Bk{\B k}

\def\BR{\B R}
\def\Bx{\B x}
\def\By{\B y}
\def\Bz{\B z}

\def\cd{\!\cdot}
\def\cl{\centerline}
\def\co{coordinate~}
\def\coo{coordinates~}

\def\di{\partial}

\def\ea{\end{eqnarray}} 
\def\ee{\end{equation}}
\def\eee{equation~}
\def\eeee{equations~}

\def\ep{\epsilon}
\def\eq{\equiv~}
\def\et{\eta}

\def\EEE{Einstein's~\eeee}

\def\fr{\frac}

 \def\ga{\gamma}

\def\ha{\frac{1}{2}~}

\def\HH{{\cal H}}

\def\inf{\infty}

\def\ka{\kappa}

\def\la {\lambda}
\def\lb{\label}

\def\lhs{left-hand side~}
\def\lll{\left(}

\def\LLL{\left[}

\def\mn{\mu\nu}

\def\na{\nabla}

\def\nn{\nonumber}
\def\nnn{\noindent}

\def\ps{\psi}

\def\ra{\rightarrow}

\def\rhs{right-hand side~}
\def\rrr{\right)}

\def\Ra{\Rightarrow}
\def\Rh{\hat{R}}
\def\RRR {\right]}

\def\si{\sigma}
\def\sq{\sqrt}

\def\sim{\simeq}

\def\su{\subsection}

\def\td{\tilde}

\def\te{\theta}

\def\ti{\times}
\def\tk{\td k}

\def\tm{\tilde m}

\def\tps{\tilde \psi}

\def\ts{\textstyle}

 \def\tV{\td V}

\def\ubm{\unboldmath}
\def\und{\underbrace}

\def\vf{\varphi }

\def\vs{\vskip 0.5 cm}

\def\wrt{with respect to~}

\def\ze{\zeta}

\def\1{{\it one}}

\def\2{{\ts{\ha}\!}}

\def\3 {\ts{\frac{1}{3}\!}}
\def\4{\ts{\fr{1}{4}\!}}
\begin{document}
\title{Toroidal Metrics: Gravitational Solenoids and Static Shells}
\author{ Donald Lynden-Bell$^{1}$\thanks{email:dlb@ast.cam.ac.uk}
\,and\,
Joseph Katz$^{1,2}$\thanks{email:
jkatz@phys.huji.ac.il}
 \\
\\  {\it$^1$  Institute of Astronomy, Madingley Road, Cambridge CB3 0HA,
United Kingdom}
\\\\
 {\it$^2$ The Racah Institute of Physics, Edmond Safra Campus, Givat Ram, 91904 Jerusalem, Israel}}

\maketitle

{\bf Abstract}
In electromagnetism a current along a wire tightly wound on a torus makes a solenoid whose magnetic field is confined within the torus.
	In Einstein's gravity we give a corresponding solution in which a current of matter moves up on the inside of a toroidal shell and down on the outside, rolling around the torus by the short way. The metric is static outside the torus but stationary inside with the gravomagnetic field confined inside the torus, running around it by the long way.
	This exact solution of Einstein's equations is found by fitting Bonnor's solution for the metric of a light beam, which gives the required  toroidal gravomagnetic field inside the torus, to the general Weyl static external metric in toroidal coordinates, which we develop. We deduce the matter tensor on the torus and find when it obeys the energy conditions.

	We also give the equipotential shells that generate the simple Bach-Weyl metric externally and find which  shells obey the energy conditions.
  \setlength{\baselineskip}{20pt plus2pt}
\section{Introduction}

We study this problem firstly to illustrate the power of thinking about the gravomagnetic field as
analogous to the magnetic field even when gravity is strong, and secondly to show that intuition
gained from studies of cylindrically symmetric space-times can often be justified. Cylindrical
systems have infinite total mass and their metrics are not flat at infinity. Often they are not  even limits of finite equilibria as some parameter tends to infinity. However, when a cylinder of finite length is bent around into circle with the ends joined to make a torus, it gives a system that can
sometimes be solved even for strong fields. When the original cylinder was rotating about its axis
the corresponding torus will be rolling about its central circle. Here we study such systems
when all the mass resides on a shell which is either a torus or in the static Bach-Weyl metric an equipotential.

	Toroidal solutions of Einstein's equations have been considered before. As early as
1922 Bach and Weyl \cite{BW} gave the simplest static solution which has been further elucidated by
Hoenselaers \cite{Ho}, and studied by Semerak et al \cite{SZZ}. The general exterior solution for the static metric in toroidal coordinates was given by Frolov {\it et al} \cite{FIU} in an investigation of cosmic strings but they did not perform the  integration that is necessary to get all the metric coefficients explicitly. Here we shall need that full solution for our external field. We also give explicit examples of equipotential shells that generate the static Bach-Weyl metric externally and find out when such shells fail to obey the energy conditions. However this paper is primarily devoted to the rolling tori which have toroidal gravomagnetic fields inside a rolling matter shell but are externally static.

The 1966 edition of the Classical theory of Fields by Landau and Lifshitz \cite{LL} gives Einstein's equations for general stationery metrics in a form that has strong analogies with Maxwell's electrodynamics. The technique identifies the points of space that lie along the time-like Killing vector so it does not extend continuously inside ergospheres where the Killing vector becomes space-like. Later  Geroch \cite{Ge} and others exploited the special properties of these equations in developments that led to the generating techniques for new solutions. We write the metric in the form 
\be
ds^2=\xi^2(dt+\AAA_kdx^k)^2-\ga_{kl}dx^kdx^l=g_{\mn}dx^{\mu}dx^{\nu},
\label{11}
\ee
where $k$ and $l$ run from 1 to 3. Since the metric is stationary $\xi,\AAA_k$ and $\gamma_{kl}$
are all independent of $t$ but in general they depend on the $x^k$. We work in the positive definite three dimensional metric of space, 
$\gamma_{kl}$. It is not a cross-section of the four metric by any surface, nevertheless we may define its Christoffel symbols $\la^m_{kl}$
and the corresponding three-dimensional Ricci tensor of this gamma space, $P^{kl}$.  We use commas to denote ordinary derivatives and semicolons to denote covariant derivatives in gamma space. The Ricci tensor of space-time will be denoted by $R_{\mu \nu}$; its indices are raised and lowered by $g_{\mu \nu}$ while the indices on $\AAA_k$ and $P^{kl}$ are raised and lowered by the gamma metric. One may show that $g^{kl}=-\gamma^{kl}, k,l=1,2,3$ and that the determinants of the metrics are related by $-g=\xi^2 \gamma $\..  In gamma-space we define the alternating tensor $\eta^{ijk} = \epsilon^{ijk}/\sq{\gamma}$ where epsilon is the alternating symbol which is unity when $i,j,k$ are an even permutation of $1,2,3$, minus one for an odd permutation and zero otherwise. $\eta_{ijk} = \sq\gamma \epsilon _{ijk}$.
The divergence and curl are defined in gamma-space by
\ba  
{ \bf div}\bE =  (1/\sq{\ga}) \di_k(\sq{\ga}E^k)~~;~~~
 ({\bf curl\,}\bE)^i= (1/\sq{\ga})\ep^{ijk}\di_jE_k~~; ~~~({ \bf grad}\, \phi)_k=\di_k{\phi};
 \lb{12}
\ea

 so ${\bf curl}~ {\bf grad}$ and ${\bf div} ~{\bf curl\,}$  are both zero. We define the gravomagnetic induction $\bBB$ by
 \be
\bBB= {\bf curl\,}\bf{\bAAA},
 \lb{13}
 \ee
  where $\AAA_k$ is the vector potential defined in the metric (\ref{11}). Clearly ${{ \bf div}\bBB }= 0$ so ${\bBB}\!$ carries the gravomagnetic flux. Landau and Lifshitz rewrite Einstein's equations in gamma space; rewriting their equations in our notation we have with $\ka = 8\pi G/c^4$,
  \be
  \xi{\bf div}\cd\!{\bf grad} \,\xi+\ha \xi^4{\bf\bBB\!\!}^2 =R_{00}=\ka(T_{00}-\ha g_{00}T).
  \lb{14}
  \ee
  Henceforth we use units with $c=1$  and $G=1$. If we now define a field intensity vector $\bHH=\xi^3\bBB$ then their second equation reads
  \be
({\bf curl\,}\bHH\!\!)^k= -2 \ka \xi T_0^k= -2\ka J^k.
  \lb{15}
  \ee
   Notice a strong resemblance of this strong field equation to Maxwell's electrodynamic equation 
 ${\bf curl\,\bH} = 4\pi {\bf j}$. In both cases the current has no divergence however in general $\bHH$ has a divergence while $\bBB$ does not. Clearly $\bHH$ is the gradient of a scalar whenever $\bf{J}$ is zero.  The $\HH^k$ are the spatial components of the twist vector $\eta^{\mu \nu \sigma \tau}\xi_{\nu}D_{\tau}\xi_{\sigma}$ where $D$ is the covariant derivative in the space-time $g_{\mu \nu}$.
 The last Einstein equation is 
 \be
 P^{kl}+ \ha\xi^2(\ga^{kl}{\BB}^2-{\BB}^k{\BB}^l)-\xi^{-1}\xi^{;k;l}=R^{kl}=\ka (T^{kl}-\ha g^{kl}T).
 \label{16}
 \ee
Einstein's equations (\ref{14}), (\ref{15}) and (\ref{16}) do not mention $\bAAA$ itself but only $\bBB= {\bf curl\,}\bAAA$, so their solution  for $\bAAA$ is arbitrary up to a spatial gradient. This gauge transformation corresponds to the transformation ${\bAAA' = \bAAA+{ \bf grad}} \,t_0(x^k)$. This corresponds to shifting the zero of time by a position-dependent function $t_0$. Under such as shift the metric remains of the same form with all the metric components independent of the new time. For our toroidal problem we want a gravomagnetic field inside the torus  to be in the toroidal direction along the unit vector ${\bf n}_{\vf}$. As the current is confined to the toroidal shell, $R^k_0 = 0$ inside, so
 ${{\bf curl}\,\bHH}=0$. Hence
\be
	\xi^3{ \bBB }= \bHH =-8I{ \bf grad}\vf,
	\lb{17}
\ee
  where $I$ is a constant which represents the total matter flux moving around of the torus by the short way. In the electromagnetic analogue the $-8$ would be replaced by a $2$ but there we have ${\bna\!\ti\!\bH} = 4\pi {\bf j}$ whereas Einstein's equation has $-16\pi$ in place of $4\pi$. We might have proceeded by inserting expression (\ref{17}) into equation (\ref{15}) and then attempt to solve it for $\xi$ within the torus where $R_{00} =0$. However empty space solutions with solely toroidal gravomagnetic fields are already available. One is obtained by boosting the Levi-Civita  solution along the its axis by a Lorentz transformation which gives a current along that axis and a gravomagnetic field around it. However that solution is unnecessarily cumbersome compared with Bonnor's \cite{Bo} beautiful exterior  metric for the gravitational field of a cylindrical light beam. In this  $ds^2$ takes the form
\be
ds^2=dt^2-(d\Bx^2+d\By^2+d\Bz^2)+(F-1)(dt-d\Bz)^2=F[dt-(1-F^{-1})d\Bz]^2-(d\Rh^2+\hat{R}^2d\B{\vf}^2+F^{-1}d\Bz^2),
\lb{18}
 \ee
 where outside the beam F is harmonic in the two dimensional $\Bx,\By$ space and
 \ba
 \Rh^2=\Bx^2+\By^2~~~;~~~F=8I\ln{(\Rh/a)}+C,
 \lb{19}
\ea
and $C$ is a constant.
For this metric the gravomagnetic intensity is
\be
\HH_k=(\xi^3 {\bf curl\,}\bAAA)_k=\Big{\{}0, F^2\Rh d_{\Rh}[(F-1)/F], 0\Big{\}}=\{0,8I,0\},
\lb{110}
\ee
so, cf.(\ref{17}), $I$ or if we like $Ic$, is the matter flux. As Bonnor shows his source for this external
solution is a cylinder of null dust travelling in the $z$ direction. It is this matter current that generates the toroidal gravomagnetic field $\bHH\!$ externally. It is this toroidal gravomagnetic field that we need INSIDE of our torus. To avoid the confusion generated by using the external part of Bonnor's metric for the inside of our torus we shall in what follows use the term Bonnor's metric to refer to his external metric. His internal metric plays no part in our calculations.
Our procedure for finding the metric of a rolling torus is to adopt a version of Bonnor's  metric  inside our torus.  We integrate equations (\ref{14}), (\ref{15}) and (\ref{16}) with delta-function contributions to $R_{00}, R_0^k$ and $R^{kl}$ on the torus, to obtain the junction conditions by which the external solution must be fitted to the internal one. We notice that $\AAA$ itself as opposed to $\BB(={\bf curl\AAA})$ is not involved in these junction conditions since none of equations (\ref{14}), (\ref{15}) and (\ref{16}) involve $\AAA$ itself. In fact the first two junction conditions are closely related to the electrodynamic ones that $\bn\times \bE$ and $\bn\,\cd\!\bB$ must be continuous where $\bn$ is the unit outward normal and the discontinuities in $\bn\,\cd\!\bE$ and $\bn\times \bB$ give surface charges and surface currents.  In the gravitational case they give a surface mass density and mass currents. The junction condition for integrating equation (\ref{16}) is a three-dimensional version of Israel's condition relating the discontinuity in the external curvatures of a three-surface to the surface stresses and currents.
Since from equation (\ref{110}) the internal metric has no gravomagnetic field penetrating the surface of the torus, discontinuities in $\bn\ti\bBB$ can all be catered for by matter currents in the surface just as they are in electromagnetism. To clarify this analogy the next section gives the flat-space solution of Maxwell's equations for a toroidal solenoid. This also serves to introduce toroidal coordinates that we use for both the electrodynamic problem and the Weyl solutions.  In our gravitational problem there is no gravomagnetic field outside the torus so the metric outside is static. Thus we can use the general solution for the static external metric in toroidal coordinates
which we develop in section 3 and illustrate with a static shell source for the Bach-Weyl metric in section 4. The main problem is then reduced that of fitting our general external solution to Bonnor's   solution which we use for the inside of our torus. This we do in section 5. In section 6 we inspect the energy conditions on the surface stresses and surface currents to ensure that they are obeyed. 
\section{The Toroidal Solenoid in Electrodynamics}
In this section we use the usual Cartesian conventions for our flat-space 3-vectors, rather than
using covariant and contravariant components.

 
 \setcounter{equation}{0}

\subsection{Toroidal coordinates}

While toroidal \coo are well known, \cite{MF} they are rarely used so we have felt it necessary to help the readers unfamiliar with them by a brief introduction. Consider two fixed points in a plane. The locus of a point having a given ratio of distances $r_1/r_2\ge1$ to those fixed points is a circle. By changing the ratio we get a set of coaxial circles. Now consider the coordinate system generated by rotating those coaxial circles about the axis that bisects the line between the two fixed points. This generates a set of coaxial tori. One \co is $\ze=\ln(r_1/r_2)\ge0$; the other axially symmetric one is $-\pi<\eta\le\pi$, the angle between the radii $r_1$ and $r_2$. The third is the angle $0\le\vf<2\pi$ about the   axis of symmetry. The cylindrical coordinates $R,  z$ can be expressed in  terms of $\ze$ and $\eta$ as follows:
\be
R=h\sinh\ze ~~~,~~~z=h\sin \eta   ~~~~{\rm with}~~~~h=\fr{a}{\cosh\ze - \cos\eta}.
\lb{21}
\ee
The flat space metric is then 
\be
dR^2+dz^2+R^2d\vf^2= h^2\lll   d\ze^2+d\eta^2+\sinh^2\ze d\vf^2    \rrr\!,
\lb{22}
\ee
Thus $h$ and $R$ are the scale factors in toroidal coordinates.
$a$ is the radius of the central line torus. On each given torus $\ze$ is constant. On the axis $R=0,\, \ze=0$ and $\ze$ is also zero at infinity. On $z=0$, $\eta= \pi$ when $R<a$ and $\eta=0$ when $R>a$.

We set
\be
S=\sq{\cosh\ze - \cos\eta}, 
\lb{23}
\ee
 then for an axially symmetric $\ps$, $\na^2\ps$ becomes in toroidal \coo
\be
\na^2\psi=\fr{S^6}{a^2\sinh\ze}\LLL  \di_\ze\lll  \fr{\sinh\ze}{S^2}\di_\ze\psi    \rrr +  \di_\eta\lll    \fr{\sinh\ze}{S^2}\di_\eta\psi      \rrr \RRR.
\lb{24}
\ee
To separate $\na^2\psi=0$ we first set $\psi=SU$, then
\be
 \di_\ze \psi=\2\fr{\sinh\ze}{S}U+S\di_\ze U,~~\di_\eta \psi=\2\fr{\sin\eta}{S}U+S\di_\eta U,~~
 \di_\ze S=\2\fr{\sinh\ze}{S},~~\di_\eta S=\2\fr{\sin\eta}{S},
\lb{25}
\ee
so
after some work,   we  find that Laplace's \eee reads
\be
\na^2\psi=\fr{S^5}{a^2}\LLL  \coth\ze\di_\ze U+\di^2_\ze U+\di^2_\eta U  + \4 \,U\RRR=0.  
\lb{26}
\ee
   It is   useful to denote
\be
u=\cosh\ze ~~~{\rm and}~~~v=\cos\eta,
\lb{27}
\ee
then \eee (\ref{26}) also reads
\be
\na^2\psi=\fr{(u-v)^{5/2}}{a^2}\Big{\{}\di_u\LLL   (u^2-1)\di_u U \RRR+ \di^2_\eta U+\4 \,U\Big{\}}=0. 
\lb{28}
\ee
Laplace's \eee separates if we write
\be
U=P(u)E(\eta)
\lb{29}
\ee
to give
\be
\fr{1}{P}\Big{\{} \LLL   (u^2-1)P'\RRR'+\4 \,P\Big{\}}=   - \fr{1 }{E}d^2_\eta E=l^2,~~~{\rm where}~~~P'=\fr{dP}{du}=d_uP.
\lb{210}
\ee
 We get thus Legendre's \eee for $P$:
\be
 \LLL   (u^2-1)P'\RRR' - L(L+1)P =0~~~{\rm with}~~~L=l-\2 ~~~~{\rm and}~~~E\propto e^{il\eta}.
\lb{211}
\ee

We note the special solution $E=E_1\eta$ when $l=0$ as well as $E=$const. We shall be interested in $u=\cosh\ze>1$ and $L$   less than an integer by $\2$, so we are interested in Legendre functions of $u$ which are not Legendre polynomials.

The general solution of the $P$  \eee  in (\ref{211}) is of the form
\be
P=a_l P_L(u)+b_l Q_L(u),
\lb{212}
\ee
but we require solutions that are finite at $u=1$ where $\ze=0$ i.e. on the axis of symmetry (and at infinity). Our potential should be symmetrical about $z=0$ so only $\cos(l\eta)$ solutions are acceptable. Hence combining the different separable solutions we find that
\be
\psi=\sq{u - v}~U,~~~~{\rm where} ~~~U=\sum_{l=0}^\inf a_l P_L(u)\cos(l\eta).
\lb{213}
 \ee
\subsection{The  Toroidal solenoid in flat space }

Here we are concerned with Maxwell's electromagnetism in flat space, not with gravomagnetism which is the subject of Sections 5 and 6. The current along the wire on the torus causes a magnetic field inside. Outside the torus there is no magnetic field so $\bna\!\ti\!\bA$ is zero there.
\begin{figure}[hp]
   \centering
   \includegraphics[width=8 cm]{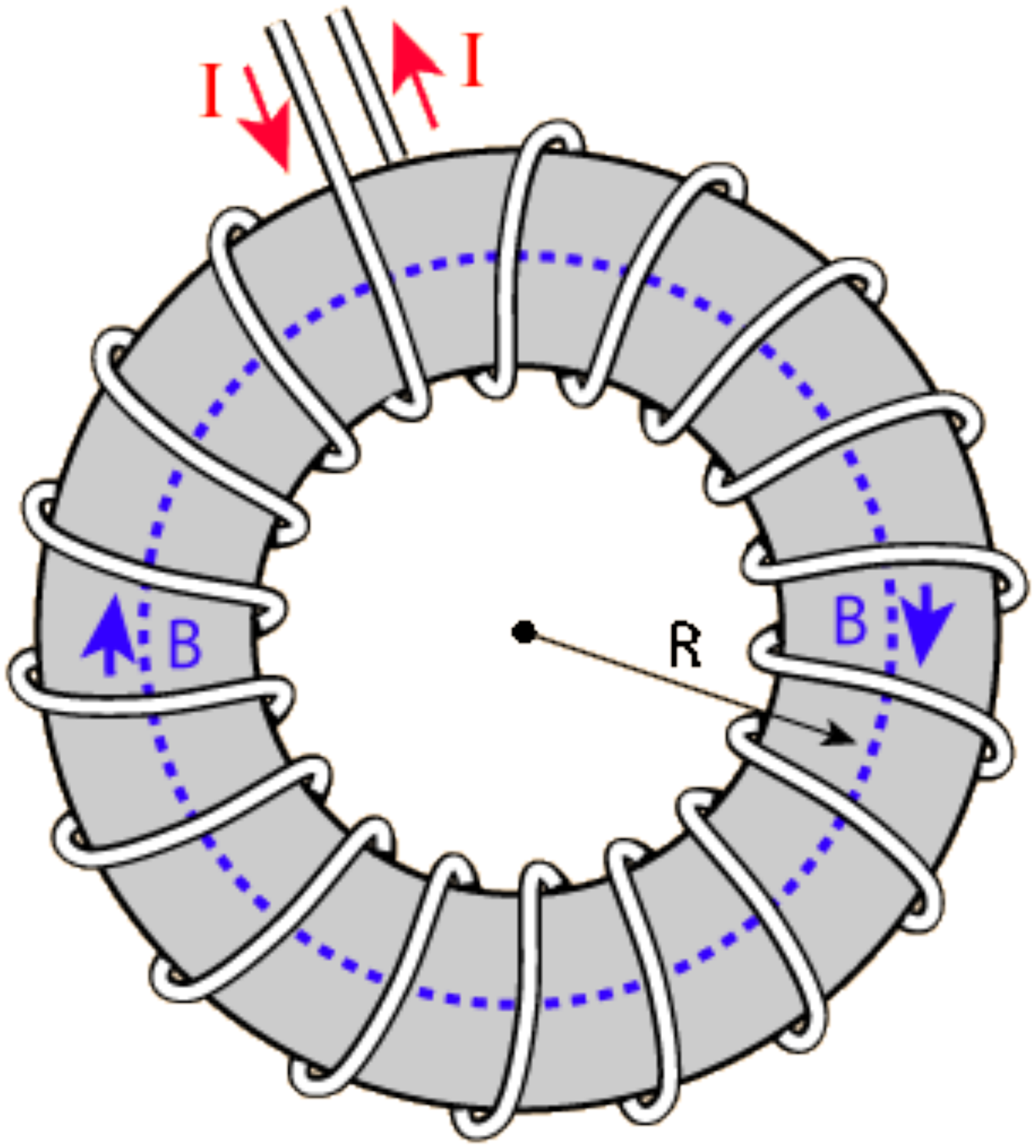} 
   \caption{\small A toroidal solenoid carrying an electric current where the windings are few and have been separated for clarity. In the mathematical problem the current is solely along $\bna{\et}$. If the solenoid had $N$ turns the total current around the pictured torus by the short way would be $NI$. This quantity is called $I$ in the text where the wire is replaced by a continuum.}
   \end{figure}

However there is a magnetic field passing along the torus so for any loop linking the torus by the short way
\be
\oint \bA\,\cd d{\bl}= \int\bB\,\cd d\bS=\Phi_T,
\lb{214}
\ee
where $\Phi_T$ is the total magnetic flux through the torus. Inside the torus the magnetic field is in the $\vf$ direction. In toroidal coordinates the local unit vectors are   
\be
\bn_\ze=h\bna\cd\ze~~~,~~~~\bn_\eta=h\bna\cd\eta~~~,~~~\bn_\vf=R\bna\cd\vf.
\lb{215}
\ee
With both $\bna\cd\bB=0$ and $\bna\!\ti\!\bB$, we may write 
\be
\bB=\fr{2I}{R}\bn_\vf.
\lb{216}
\ee
$I$ is a constant current. To find a vector potential $\bA$ which gives this $\bB$ we write
\be
\bA=A\bna\eta, 
\lb{217}
\ee
$\bna\!\ti\!\bA$ will be in the $\vf$ direction if $A$ is a function of $\ze, \eta$ or $u, v$ only. The condition that $\bna\!\ti\!\bA=\bB$ is easily found to be
\be
\fr{\di A}{\di \ze}= \fr{2Ih}{ \sinh\ze}.
\lb{218}
\ee
$A$ is thus an integral over $\ze$. The arbitrary function of  $\eta$   is determined since $A$ must be zero on the line torus where $\ze\ra\inf$. The integral is readily evaluated by writing $x=e^\ze$ to give
\be
A=\fr{2Ia}{1-v^2}\ln\LLL   \fr{x^2-1}{x^2-2vx+1}   \lll\fr{x-1}{x+1} \rrr^{\!\!v}    \RRR=\fr{2Ia}{\sin^2\eta}\ln\bigg{\{}    \fr{\sinh\ze}{\cosh\ze-\cos\eta} \LLL\tanh(\2\ze)\RRR^{\cos\eta}       \bigg{\}} \!.
\lb{219}
\ee
Despite appearances this expression is regular as $\eta\ra0 ~{\rm and}~  \pi$, which is not readily seen from the expression in terms of $x$.

Outside the solenoid $\bA$ has to be a gradient since the magnetic field is zero. This is readily accomplished by taking $A$ to be the function of $\eta$ that is achieved on the solenoid itself where $\ze=\ze_s$ or $x=x_s=e^{\ze_s}$. Thus outside
\be
A=\fr{2Ia}{\sin^2\eta}\ln\LLL   \fr{x_s^2-1}{x_s^2-2\cos\eta \,x_s+1}   \lll\fr{x_s-1}{x_s+1} \rrr^{\!\!\cos\eta}    \RRR\!.
\lb{220}
\ee
For $\eta\ra\pi$,
\be
A\stackrel{\eta\ra\pi}{\longrightarrow}2Ia\LLL    \ln\sq{\fr{x_s-1}{x_s+1}}   - \fr{x_s}{(x_s+1)^2}  \RRR\!.
\lb{221}
\ee
The surface current in the $\bn_\eta$ direction around the torus is given in terms of the discontinuity of $\bB$ so
\be
4\pi\bJ=-\fr{2I}{R}\bn_\eta= - \fr{2I}{\sinh\ze_s}\bna\eta.
\lb{222}
\ee
This completes the solution for the magnetic field of a toroidal solenoid. It is a close analogue of the gravomagnetic field discussed in Section 5 and 6.

\section{The general static Weyl metric in toroidal coordinates}
  \setcounter{equation}{0}
Weyl takes the metric in the form 
\be
e^{-2\psi}dt^2 - e^{2\psi}\LLL   e^{2k} \lll    dz^2 +  dR^2       \rrr  +R^2d\vf^2 \RRR= e^{-2\psi}dt^2 - e^{2\psi}\LLL   e^{2k}h^2 \lll   d\ze^2+d\eta^2   \rrr  +R^2d\vf^2 \RRR\!.       
\lb{31}      
\ee
Then, in empty axially symmetric spaces Einstein's \eeee give $\na^2\psi=0$ where $\na^2$ is the flat space operator. Also setting
\be
D=\di_R -i\di_z,
\lb{32}
\ee
we have the Weyl \eeee
\be
Dk=\4 Re^{4\psi}De^{-2\psi} De^{-2\psi}.   ~~~{\rm So}~~~DkD\ln R=(D\psi)^2.
\lb{33}
\ee
From (\ref{31}) we see that $\zeta+i\eta$ may be obtained by conformal transformation of $z+iR$
so if we set 
\be
D_*=\di_\ze + i\di_\eta,
\lb{34}
\ee
  \eee  (\ref{33}) implies via conformal transformation  
\be
D_*kD_*\ln R=(D_*\psi)^2=(\di_\ze\psi)^2 - (\di_\eta\psi)^2 + 2i\di_\ze\psi\di_\eta\psi.
\lb{35}
\ee
Now
\be
D_*\ln R=D_*\LLL    \ln\sinh\ze - \ln(\cosh\ze - \cos\eta) \RRR= \coth\ze - S^{-2}(\sinh\ze + i \sin\eta).
\lb{36}
\ee
Hence our \eeee for $k$ become
\ba
&& \di_\ze k\! \lll   \! \coth\ze - \!\fr{\sinh\ze}{S^2} \!\!  \rrr +\di_\eta  k\fr{\sin\eta}{S^2} = (\di_\ze\psi)^2 - (\di_\eta\psi)^2\nn,
\\
&&\di_\ze k \lll \!\! \fr{\sin\eta }{S^2}\!\rrr - \di_\eta k\lll \! \coth\ze\! -\!\fr{\sinh\ze }{S^2}\! \rrr\!=\! -2\di_\ze\psi\di_\eta\psi.
\lb{37}
\ea
For regularity we need $k=0$ on axis where $\ze=0$.  So $k$ may be found by integrating $\di_\ze k$ from $0$ to $\ze$ at constant $\eta$. Therefore we eliminate $\di_\eta k$ and obtain after a simple calculation  in terms of $u, v$ defined in (\ref{27})
\be
\di_u k=\lll   u - \fr{u^2-1}{S^2}  \rrr \LLL  (u^2-1)(\di_u\psi)^2 -(1-v^2)(\di_v\psi)^2   \RRR+\fr{2}{S^2}(u^2-1)(1 - v^2)\di_u\psi\di_v\psi~~~.
\lb{38}
\ee
However our potentials are all of the form $\psi=SU$ so putting this form into (\ref{38}) and simplifying,
\ba
&&\di_u k= - \4 (1+uv)U^2 - \LLL   v(u^2-1)\di_u U-u(1-v^2)\di_v U \RRR U\nn~~~~~~~~~~~~~~~~~~~~~~~~~~~~~~~~~~
\\
&&~~~~ +(1-uv)\LLL     (u^2-1)(\di_uU)^2   -(1-v^2)(\di_v U)^2\RRR+2(u^2-1)(1-v^2)\di_uU\di_v U.
\lb{39}
\ea 
 The general solution for $U$ involves the sum  given in (2.13). The general solution for $k$ must be obtained by integrating (3.9) which  is quadratic in $U$ so that leads to double sums.    We write 
\be
L=l-\2~~~,~~~M=m-\2~~~{\rm and}~~~c(n)=\cos(n\eta).
\lb{310}
\ee
and after performing the $v$-differentiations the integral for $k$ is given in terms of
 $P_L$ and $P_M$,    Legendre functions of $u$: 
\ba
&&\!\!\!\! k= \sum_{l=0}^{\inf}\sum_{m=0}^{\inf}a_la_m\!\int_1^u\!\!\! \!\! \LLL P'_LP'_M(u^2\!-\!1)(1\!-\!uv)\!- \!\4 P_LP_M(1\!+\!uv)  \! - \! \2(P_LP_M)'(u^2\!-\!1)v \RRR du~c(l)c(m)
\nn
\\
&&+ \sum_{l=0}^{\inf}\sum_{m=0}^{\inf}a_la_m\!\!\int_1^u\!\!\!  \!\!\LLL \2 P'_LP_M m(u^2\!-\!1)\!+\!\!\ts{\fr{1}{8}}P_LP_M(l\!+\! m)u     \RRR du\LLL c(l\!+\!m\!+\!1) \!- \!c(l\!+\!m-\!1)         \RRR\nn
\\
&&+ \sum_{l=0}^{\inf}\sum_{m=0}^{\inf}a_la_m\!\!\int_1^u\!\!\!  \!\!\LLL  \!-~\!\2 P'_LP_M m(u^2\!-\!1)\!+\!\ts{\fr{1}{8}}P_LP_M(l\!-\! m)u     \RRR du\LLL  c(l\!-\!m\!+\!1) \!- \!c(l\!-\!m-\!1)         \RRR\nn
\\
&&+\sum_{l=0}^{\inf}\sum_{m=0}^{\inf}a_la_m\!\!\int_1^u\,\2 \!P_LP_Mlm(1-uv) \,du\LLL   c(l+m) - c(l-m)  \RRR\!.
\lb{311}
\ea
In Appendix A we show how the indefinite integrals in (\ref{311}) can be evaluated in terms of Legendre functions. This results in $k$ being given in the form 
\ba
&&k={\ts\fr{1}{8}} \sum_{l=0}^{\inf}\sum_{m=0}^{\inf}a_la_m\LLL  c(l+m+1)k^1_{l,m} +c(l+m)k^0_{l,m}+c(l+m-1)k^{-1}_{l,m}                 \RRR\nn~~~~~~~~
\\
&&~~+{\ts\fr{1}{8}} \sum_{l=0}^{\inf}\sum_{m=0}^{\inf}a_la_m\LLL  c(l - m+1)k^{1}_{l,-m} +c(l-m)k^0_{l,-m}+c(l-m-1)k^{-1}_{l,-m}               \RRR\!,~~~~~~~~~~
\lb{312}
\ea
where $k^n_{l,m}$ are known terms of Legendre functions themselves and their derivatives.
Thus $\psi$ of the form given in  (\ref{213}) and $k$ of the form given in (\ref{312}) and the $k^n_{l,m}$ given in Appendix A constitute the general equatorially-symmetric Weyl solution of \EEE in toroidal \coo$ \!$.
All finite sum solutions are singular on the line toroid at $R=a, z=0$. We use this general solution
in sections 5 and 6 where we discuss the gravitational solenoid but before tackling that more complex problem we consider the simplest static toroidal shell source. This sheds light on the conicity which is a common feature of both problems but is most clearly illustrated without
the complications of  moving sources.


\section{The Bach-Weyl metric generated by a static shell}

 \setcounter{equation}{0}
 

\su{The model}

The simplest of the toroidal solutions (\ref{213}) is given by keeping only $a_0$ non-zero. Then
\be
U=a_0 P_{-1/2}(u)\eq a_0 P,
\lb{41}
\ee
which is independent of $\eta$. Henceforward we shall drop the subscript $-1/2$ and merely write $P(u)$ on the understanding that this $P$ is the Legendre function of order $-1/2$.  
Substituting expression (\ref{41}) into (\ref{39}) all the $v$-derivatives vanish and we obtain an expression for $\partial k/\partial u $, which integrates by parts. Using the recurrence relation
for Legendre functions the final integral can be put in terms of $P$ and $P_{1/2}=P_{-3/2}$. The solution for $k$ with the boundary condition that $k=0$ on axis is then
\be
k=\4 \,a_0^2{\Big\{}2P(P_{\ha}-uP)+v\LLL P^2-(P_{\ha\!\!})^2\RRR{\Big\}}
\lb{42}
\ee
Expression (\ref{45})  for $k$ coupled with
\be
\psi=a_0\sq{u-v}P,
\lb{43}
\ee
constitute the complete Bach-Weyl metric. If we consider these as solutions everywhere then they are singular on the line toroid that forms the circle $R=a, z=0$ where $u$ is infinite. Indeed using \cite{OLBC} formula 14.8.14 we find
\be
P(u) \stackrel{u\ra\inf}{\longrightarrow}\fr{\sq{2}}{\pi u^{1/2}}\ln(8u)
\lb{44}
\ee
Thus, for large    $u$
\be
\psi\stackrel{u\ra\inf}{\longrightarrow}\fr{a_0\sq{2}}{\pi   }\ln(8u)
\lb{45}
\ee
and at large fixed $u$, 
\be
(R-a)^2+z^2\sim\fr{a^2}{u^2},
\lb{46}
\ee
so $u$ is then constant on a torus of small radius $a/u=s$ about the singular circle. The behaviour  of $\psi$, see (\ref{45}), then shows us, by comparison with both the classical result and that of the cylindrical line source in relativity, that the mass per unit length is $\sq{2a_0}=2\pi\mu$. 
The metric will therefore suffer from the problems of line sources when the singular line is approached too closely. However, when the Bach-Weyl solution is generated as the
external metric of a massive equipotential shell that obeys the energy conditions, that external part of the metric can not be an unphysical one. The asymptotic form of $\psi$ at large $r$  gives the total mass $m$ of the system. $\psi\ra\fr{\sq{2}a_0a}{r}$ so the mass is $\tm=\fr{m}{a}=\fr{GM}{ac^2}=\sq{2}a_0=2\pi \mu.$
While this sounds just what we might expect, the proper length of the singular line is $e^{2\psi}2\pi a\ra\inf$ since $U\ra a_0$ and $u^{1/2}\ra\inf$.
We shall take a massive shell that lies on an equipotential surface $\psi=V=$ const  of the Bach-Weyl solution. This surface is a toroid  in that it has the topology of a torus but, as seen in  Figure 2, it lacks the circular small cross-section of a true torus of constant $u$.

\begin{figure}[hp].
   \centering
   \includegraphics[width=10 cm]{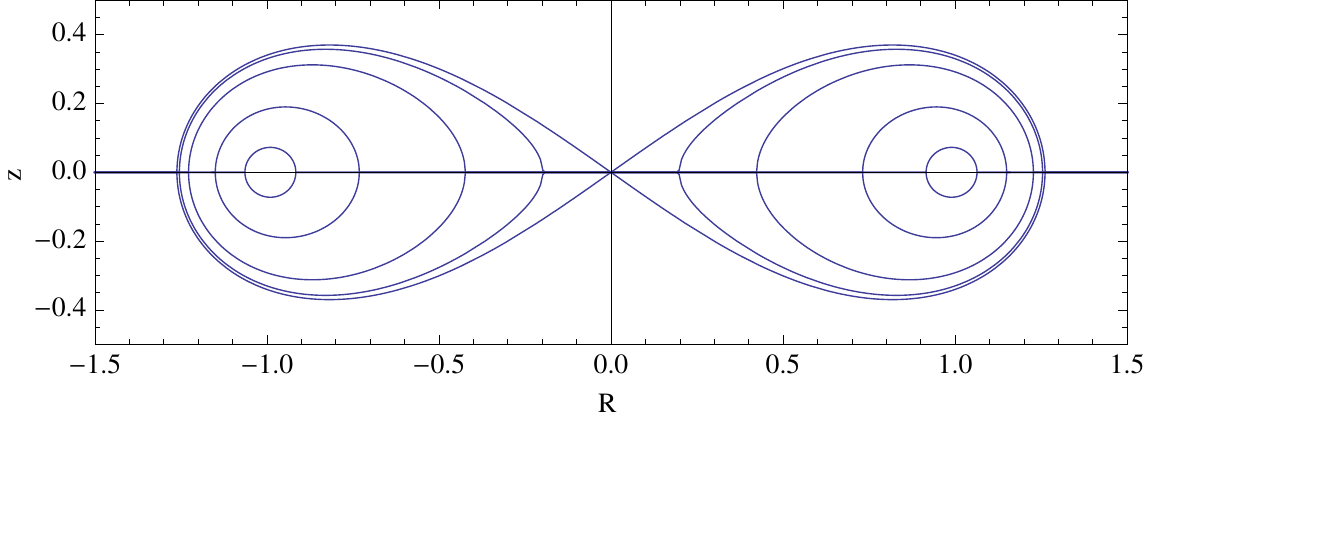} 
   \caption{\small  Equipotentials in $R,z$ coordinates for $\td\psi=1$ the largest toroid   crossing the origin, through $\td\psi=1.01, 1.05, 1.2$ and $\td\psi=1.5$ the smallest one almost circular.We choose one of these surfaces and make it a massive shell with constant internal potential. There is a limit to the mass that can be put on it given in Figure 5. }
   \end{figure} 
  Inside our shell there is no matter, $\na^2\psi=0$ and as $\psi$ is constant on the boundary it is $V$ everywhere. The equation for $k$ then shows it must be  a constant $\B k$ everywhere inside the toroid and space-time is {\it locally flat} :
 \be
 d\B s^2= \B g_{\mn}d\B x^\mu d\B x^\nu=e^{-2V}dt^2 - e^{2V}\LLL e^{2\B k}(d\B z^2+dR^2) +R^2d\vf^2\RRR\!.
 \lb{47}
 \ee
 Space-time is not globally flat because the axis of symmetry does not touch the locally flat space within the toroid, so we cannot use the regularity on the axis to show that $\B k=0$. We determine  $\B k$ below.
 
The \eee of the toroid as seen from outside  is, say, $z= z(R)$, defined parametrically by (\ref{21}), and, following (\ref{46}), $z(R)$ is defined by the condition that
\be
\tV=\fr{V}{\tm}=\sq{\2(u-v)}P(u)={\rm Const}~~~{\rm where}~~~\tm=\fr{m}{a}.
\lb{48}
\ee

\su{ The junction condition}

  The metric on the shell
  is thus, according to (\ref{31}), with $dz=z_RdR$,  
  \ba
&&  d \si^2=h_{ab}d\te^ad\te^b=e^{-2V}dt^2 - f^2(1+z_R^2)dR^2  - e^{2V}R^2d\vf^2 ~~~,~~~\te^0=t, \te^1=R, \te^2=\vf,\nn
\\
&&~~~f=e^{V+k}~~~~{\rm and}~~~ z_R=(\di_R z)_\psi= - \fr{\di_R\psi}{\di_z\psi}=\fr{\sq{u^2-1}}{\sin\eta} \LLL  \fr{ -\2  vP+(1-uv) P' }{\2 uP+(u^2-1)P' }\RRR\!.
\lb{49}
 \ea
 
 Let $\B z=\B z(R)$ be the \eee of the toroid as seen from inside. Accordingly, see (\ref{47}),
  its metric is
 \be
 d\B \si^2=\B h_{ab}d\te^ad\te^b=e^{-2V}dt^2 - \B f^2(1+\B z_R^2)dR^2  - e^{2V}R^2d\vf^2~~{\rm where}~~\B z_R=\fr{d\B z}{dR}~~{\rm and} ~~\B f=e^{V+\B k}.
 \lb{410}
\ee 
Since  (\ref{49})  and  (\ref{410}) represent the same hypersurface we must have 
\be
e^{2k}(1+z_R^2)  = e^{2\B k}(1+\B z_R^2).
\lb{411}
\ee
This junction condition gives the differential \eee for the contour $\B z(R)$ as seen from within the toroid which may be written
\be
{\B z}^2_R=z_R^2 +\lll   e^{2(k-\B k)}  -1   \rrr(1+z^2_R) .
\lb{412}
\ee 
For given $\tV$ and $\tm$ both $k(u)$  and $z(u)$   are known on the toroid but $\B z(u)$ and the constant $\B k$ are as yet unknown. Explicitly $k$ is given by (\ref{42}) but with $v$ given as a function of $u$ and $\tV$ by (\ref{48}).
It is useful in what follows to set
\be
e^\chi=\sq{1+z_R^2}~~~,~~~e^{\B\chi}=\sq{1+\B z_R^2}.
\lb{413}
\ee
In terms of $\chi$ and $\B \chi$ the junction condition (\ref{411}) may be written 
 \be
fe^\chi=\B fe^{\B\chi}~~~{\rm or}~~~\B k+ \B \chi= k+\chi.  
\lb{414}
\ee

\su{Evaluation of the ``conicity" $\B k$}

   $\B k$ characterizes the conicity within the toroid. The conicity, defined as in \cite{BLSZ}, is    the circumference of a circle of radius $(R+dR)$ minus the circumference of a circle of radius $R$, that is $2\pi e^{V}dR$  divided by $2\pi$ times the proper distance between the circles,  that is
 \be
 \fr{2\pi e^VdR}{2\pi e^{V+\B k}dR}=e^{- \B k}.
 \lb{415}
 \ee
 Thus if $\B k>0$ there is a deficit angle, a local ``conicity", typical of  a circle on a cone centered on its apex or of spacetimes with a string. In the present case as we shall see there is   ``anti-conicity", i.e.  $\B k<0$.
 
 To calculate $\B k$ we look at the point $u_0$ where $\B z(R)$ reaches a maximum for a given $V$, that is where $\B z_R^2=0$.  
 Differentiating (\ref{412}) \wrt $u$ we find, writing a dash for derivatives \wrt $u$ 
\be
\B z_R\B z'_R= e^{2(k - \B k)}\LLL k'(1+z^2_R)+z_Rz'_R\RRR,
\lb{416}
\ee
so where $\B z_R=0$, 
\be
k'= - \fr{z_Rz'_R}{(1+z^2_R)}.
\lb{417}
\ee
Since $k(u)$ and $z_R(u)$ are known, this \eee may be used to determine the value of $u$, $u_0$ where $\B z_R=0$.
We have here two conditions, one  (\ref{417})  that fixes $u_0$ and one from (\ref{414}) or (\ref{412}) that gives us $\B k$. Thus where $u=u_0$, differentiating (\ref{414}) where needed,
 \be
\B\chi=\B\chi'=0~~~\Ra~~~ \B k= k+\chi ~~~{\rm and}~~~k'+\chi'=0.
 \lb{418}
 \ee
From (\ref{48}) we find that on the toroid
 \be
 v(u,\tV)=u - \fr{2\tV^2}{P^2(u)},
 \lb{419}
 \ee
 and with $v$ understood to be this function, we have, following (\ref{42}),
 \be
 k=\tm^2 \td k(u, \td V)~~~{\rm where}~~~ \tk(u,\tV)=P(P_{\ha}-uP)+\2v\LLL P^2-(P_{\ha\!\!})^2\RRR.
 \lb{420}
 \ee
 Instead of seeking the values of $\B k$ and $u_0$ where $\B z(R)$ maximizes on a chosen potential surface $V$, we find it convenient to choose values of $u_0$ and $\tV$ and to seek   $\B k$ and $\tm$ for which those values obey the maximizing condition (\ref{418}). Using (\ref{420}) we   thus obtain $\tm$ and $\B k$ as functions of $u_0$ and $\tV$ from (\ref{418}):
 \be
\B k(u_0, \tV)=\tm^2\tk+\chi~~~{\rm with}~~~ \tm(u_0, \tV)= \sq{\fr{-\chi'}{\tk'}}.
 \lb{421}
 \ee
\begin{figure}[tbp].
\includegraphics[width=8 cm]{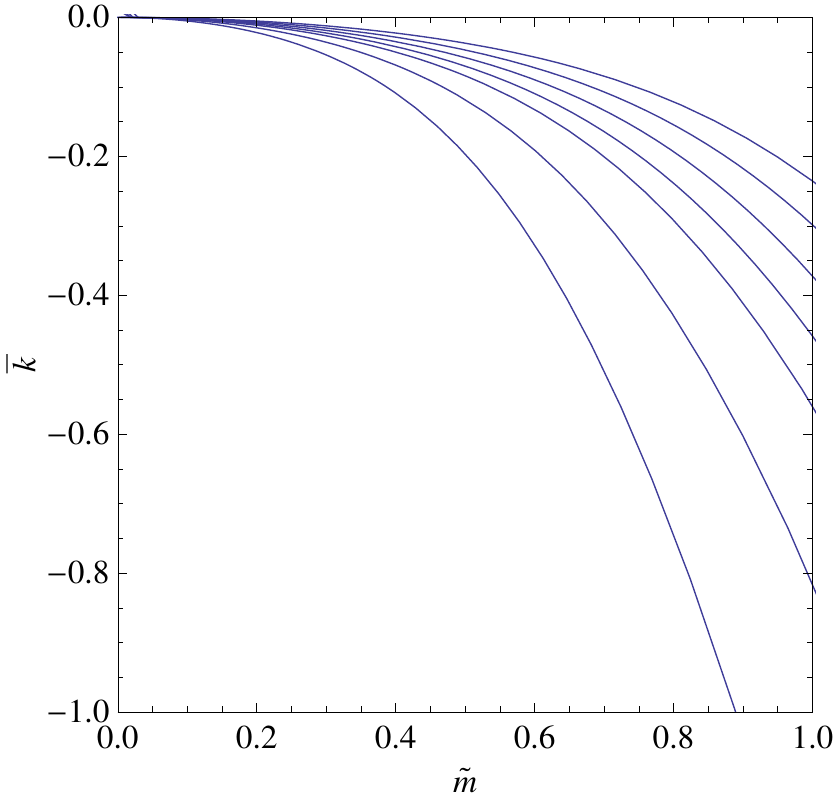} 
\caption{\small  $\B k$ as a function of $\tm$ for, from top to bottom, $\tV= 1.05, 1.10, 1.15, 1.20, 1.25, 1.35, 1.5$. }
\end{figure}
  We used Mathematica to make a parametric plot of $\B k$ as a function of $\tm$ for different values of $\tV$. We see in Figure 3 that $\B k$ is negative. For a given   $\tm=2\pi\mu$ , $\B k$ becomes more negative with increasing $\tV$. 
 
 \subsection{ Surface pressures and surface mass-density on static toroidal shells}

We calculated the surface density and the surface stresses in the shell using the space-time
fitting conditions of Israel. 
For a general position on the toroid these expressions prove tediously long and unenlightening when all the differentiations are inserted. They become somewhat simpler when $v=\ep_v=\pm1$ or $\sin\eta=0$ i.e. on the equatorial plane where they reach extreme values. 
 We obtain expressions for the energy per unit area and the tensions or pressures in the toroid:
\ba
&& \ka\si= \lll  \fr{1}{\B f}  - \fr{1}{f}    \rrr\lll  \fr{\ep_v}{R} - \B f\di_z n^z     \rrr -\fr{\ep_v}{f}\di_R(2\psi+k)\nn
\\
&&
\ka p_\et= -\fr{\ep_v}{R}\lll  \fr{1}{\B f}  - \fr{1}{f}    \rrr\nn
\\
&&
\ka p_\vf= \lll  \fr{1}{\B f}  - \fr{1}{f}    \rrr\B f\di_z n^z + \fr{\ep_v}{f}\di_R k
\lb{422} 
\ea
Notice that if one sets $f=\B f=k=1$, the mass-energy density per unit length reduces to its classical value
\be
\ka\si= - 2v\di_R\psi.
\lb{423}
\ee
$\si$, $p_\et$ and $p_\vf$ are quite complicated parametrized functions of $\tm$ and $\tV$.   One possible test that the results are sensible consists in calculating the equilibrium of the forces in the equatorial plane in the limit of small $\tm$.   $R$ is the radius of curvature in the $\vf$ direction in the classical limit. Let $b$ be the curvature radius in the $\et$ direction. The non-relativistic equilibrium of forces at $v=\pm1$ is given by
\be
\ep_v\fr{p_\vf}{R}+\fr{p_\et}{|b|}={\4}\ka\si^2~~~\Ra~~~\fr{(\ep_vp_\vf/R)+(p_\et/|b|)}{ \ka\si^2/4}=1
\lb{424}
\ee
Now,
\be
\fr{1}{b}=\fr{ (u-\ep_v)}{\sq{u^2-1}}~~~~{\rm and}~~~~\fr{1}{R}=\fr{(u-\ep_v)}{\LLL 1+2(u-\ep_v)\fr{P'}{P}\RRR} -\ep_v\sq{(u^2-1)}.
\lb{425}
\ee

We calculated with Mathematica the ratio in (\ref{424}) for $\tV=1.2$ and a mass   $\tm\sim0.01$. We found that for $v=+1$, the ratio is $\sim 1.01$ and for  $v=-1$, it is $\sim1.02$. These are reasonably close to $1$. For a mass $\tm\sim0.2$ which is rather relativistic the ratios are respectively $\sim 1.2$ and $\sim1.6$. 
\begin{figure}[ht].

   \includegraphics[width=10 cm]{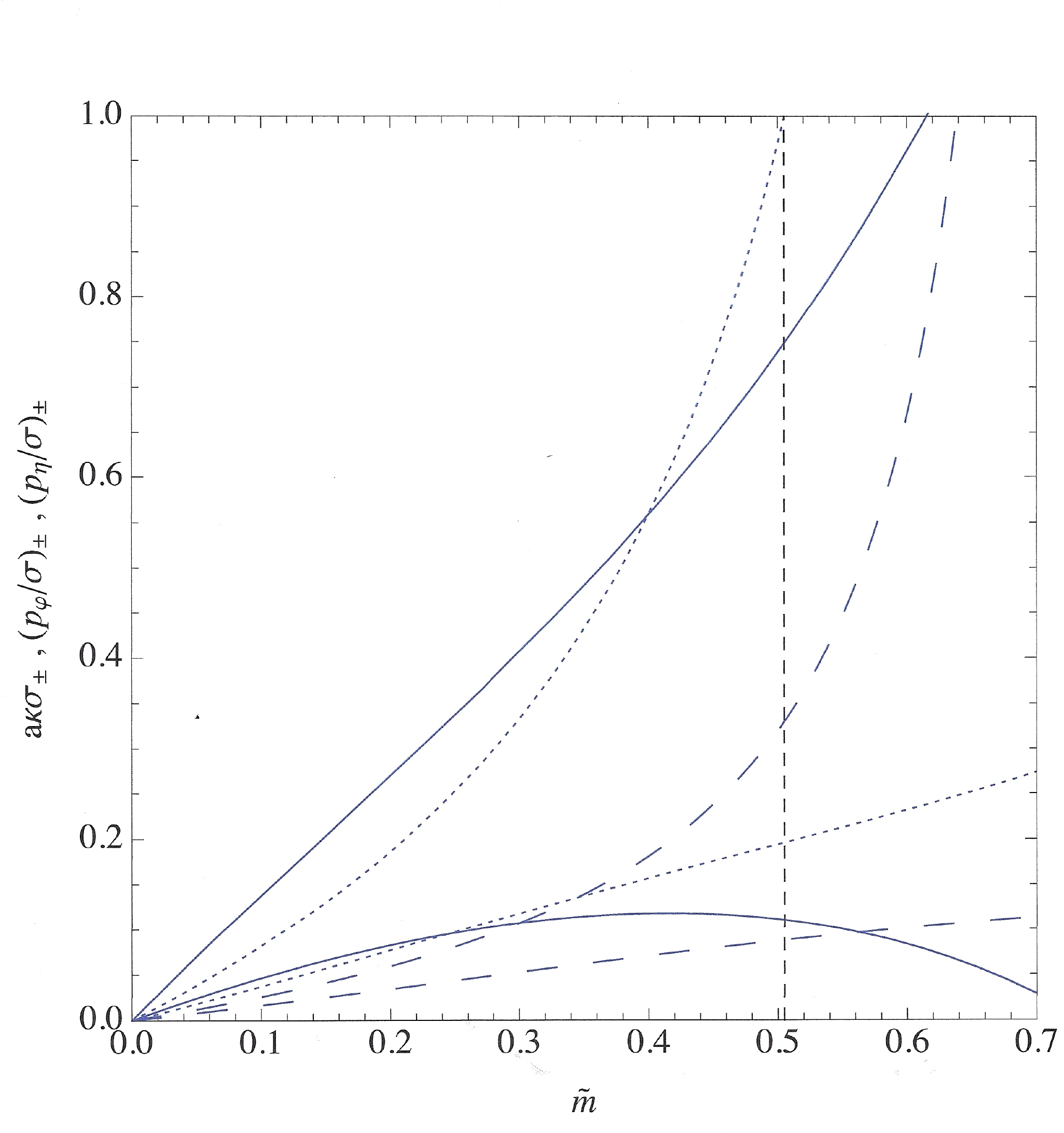} 
    
  {\small  Figure 4: The vertical dashed line is the limit where the energy condition $(p_\vf/\si)_-\!\le1$ breaks down. Plain line: the matter densities or $a\ka\si_{\pm}$ as   functions of $\tm$ for $\tV=1.2$. $\si_-$ is the lower curve.    \nnn  Thick dashed lines: $(p_\et/\si)_\pm$ as   functions of $\tm$ for $\tV=1.2$. The lower line is for $v=1$. The $p_\et$ pressure is never too great. Tiny dotted lines:  $(p_\vf/\si)_\pm$ as   functions of $\tm$ for $\tV=1.2$. The lower line is for $v=1$. For $\tm=0.505$   the pressure $p_{\vf}/\si=1$.    }
   \end{figure} 
$\si_\pm$ varies on both sides of the toroid in the equatorial plane in terms of the mass $m$. The characteristic property of   $\si_-$ is that it becomes negative for $\tm\gtrsim0.75$ beyond which the solution becomes unphysical.  Figure 4
 represents the ratio $(p_\vf/\si)_\pm$ and    can see  that the pressure in the inside of the toroid, $v=-1$ or $\et=\pi$ becomes too big for $\tm\gtrsim 0.505$.   

\begin{figure}[ht].
 
   \includegraphics[width=10cm]{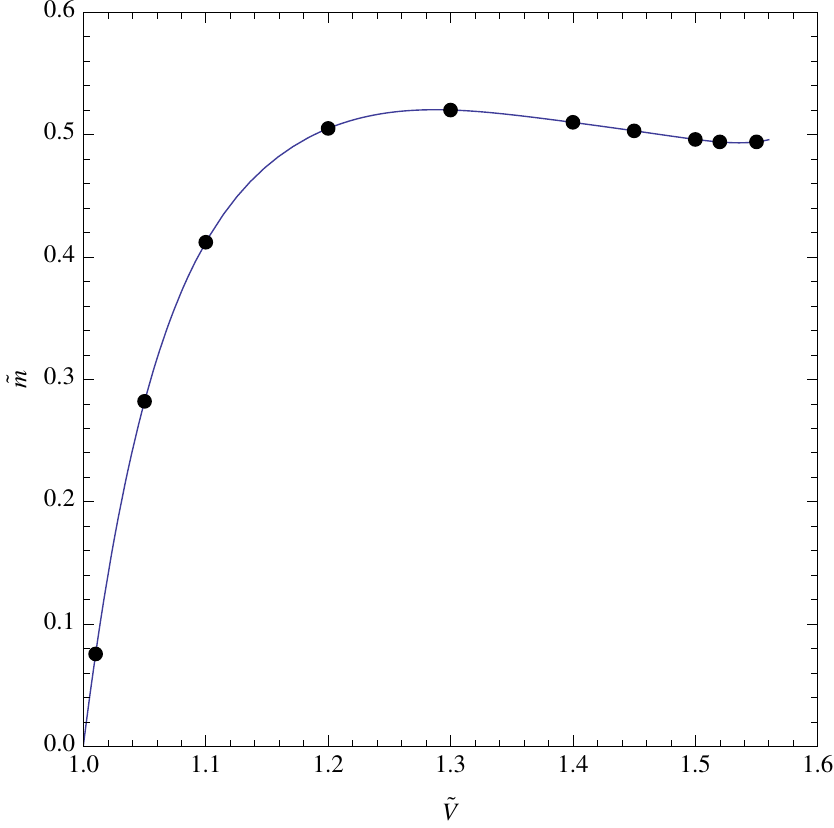}  
  
  {\small Figure 5: The dominant energy condition limits the mass that can be put on the equipotential $\tps=\tV=\fr{V}{\tm}$. This maximum mass is plotted as a function of $\tV$. The limiting condition is $(\fr{p_\vf}{\si})_-=1$. For the shape of the equipotentials see Figure 2.}
   \end{figure} 
 Figure 5 shows that  for $\tm\lesssim 0.505$ the pressure in the $\et$-direction is never too large. Thus for $\tV=1.2$ all energy conditions hold for $\tm$ not too large: $0<\tm\lesssim0.505$. Figure 5 represents the limits   of $\tm$ as a function of $\tV$. The line is a polynomial interpolation. The toroids do not satisfy the energy condition above  that line.

Let $\si _\pm$, $p_{R\pm}$ and $p_{\vf\pm}$ represent the energy densities and pressures in the equatorial plane at $v=\pm1$ where we expect to find limiting conditions. The dominant energy  conditions are $\si_{\pm}\ge0$, $-1\le(p_\et /\si)_{\pm}\le1$ and $-1\le(p_\vf /\si)_{\pm}\le1$. 
We shall illustrate the situation with numerical results which are typical of the general situation taking $\tV=1.2$. 

\section{A toroidal solenoid's metric  and  junction  conditions}
 \setcounter{equation}{0}
 In Section 3 we gave the general equatorially-symmetric static Weyl solution of Einstein's equations in toroidal coordinates. Our aim now is to fit this exterior solution to Bonnor's metric which has a toroidal gravomagnetic field. We fit on a torus of the exterior Weyl solution. This Weyl metric can be written with $\xi=e^{-\psi}$.
At large distances $\et\ra0$ and $u\ra1, \sq{u-v}\ra\sq{2 }a/r$. Hence
 $\psi\ra\sq2(a/r)\sum a_l=M/r$.
Notice that all the $a_l$ contribute to the asymptotic mass $M$, not just $a_0$. We use the external potential (\ref{213}) outside our toroidal shell where $\zeta <Z$ that is $u<u_s=\cosh Z$. 
 The same conicity problem occurs when we try to fit the primitive form of Bonnor's metric inside our torus.
   We therefore generalise Bonnor's metric to include an extra constant conicity term $e^{2\Bk}$;  this is easily done since Bonnor's metric has a Killing vector and  Einstein's differential equations are local. We may replace  $\B{\vf}$ in Bonnor's solution by any constant multiple of it and the metric will still be a solution locally.  
   Of course the metric no longer satisfies the condition that it is regular on the axis. It will now 
   have a string discontinuity there, but the axis is not included in the interior of our torus, so no such  discontinuity occurs in the part of the space we use. Replacing $\B{\vf}$ by 
   $e^{\Bk} \vf$ in  Bonnor's metric (\ref{18})
   and setting $\BR =\Rh e^{-\Bk }$ we obtain the metric 
 \be
 ds^2=F[dt-(1-F^{-1})d\Bz]^2- (e^{2\Bk}d\BR^2+\BR^2d\vf^2+F^{-1}d\Bz^2).
 \lb{51}
 \ee
This is the one we use for the interior of our torus. Rewriting $F$ as a function of $\BR$ rather than  $\Rh$ cf (\ref{19}) ,$F=8I~\ln{(\BR/a)}+\BC$
where $\BC$ is another constant. To make easy comparison with our work on equipotential toroids 
in Section 4, we choose $\BC=\exp[-m\sqrt{2u_s}P(u_s)]$ so that the potential at $R=a$ is that found earlier for the Bach-Weyl toroids. This formula defines our parameter $m$ which is no longer the total mass which we call $M$. However for large $\tilde{V}$ the Bach-Weyl toroids approximate tori so for them $I=0,m=M$. Because of the change of metric from (\ref{18}) to (\ref{51}) we now have different field components,~~$\HH_{\vf}=8e^{-\Bk}I$ rather than $8I$ cf (\ref{110}). The constant $\Bk$ is determined by the same procedure as we used for static toroids.
\subsection{Fitting the potentials and the  gamma metrics} 

The internal metric (\ref{51}) and the external metric (\ref{31}) with the potential given by (\ref{213}) and
$k$ by (\ref{312}), must give the same induced metric on the torus. Comparing the coefficients of $dt^2$ we find $F=e^{-2\psi}$ hence from (\ref{213}) we have
\be
\sum a_lP_L(u_s)\cos(l\et)=-\ha\ln [F(\BR)]/\sqrt{u_s-\cos \et}.
\lb{52}
\ee
Comparing coefficients of $d\vf^2$ we find
\be
\BR^2=R^2e^{2\psi}=R^2/F(\BR)=a^2(u_s^2-1)/[(u_s-\cos \et)^2 F],
\lb{53}
\ee
which give $\BR$ as an implicit function of $\et$ on the torus. Thus $F(\BR)$ in (\ref{51}) is a 'known'  function of $\et$, and the coefficients in the Fourier series of the \rhs of (\ref{52}) divided by $P_L(u_s)$ appropriately give us the $a_l$. With those known the function $k$ is known from the sum (\ref{312}). Comparing coefficients of $ d\et^2$ we find, on re-ordering the terms and multiplying by F
\be
\Bz'^2=e^{2k}h^2-Fe^{2\Bk}\BR'^2,
\lb{54}
\ee
where a $'$ stands for a derivative with respect to $\et$.
The unknowns are the function $\Bz(\et)$ that gives $\Bz$ as a function of $\et$ on the torus and the constant $\Bk ;~~\BR(\et)$ is given as the solution to (\ref{53}). The boundary condition is that $\Bz=0$ at $\et=0$. We must still determine $\Bk$; for any selected value of it we can imagine integrating (\ref{54}) starting from $\et=0$. Eventually the \rhs will reach zero where $\Bz$ reaches its maximum. However in general d$_\et$(\rhs$\!\!$) will not be zero
when \rhs$=0$. But $d_\et(d_\et\Bz)^2=2d_\et\Bz d^2_\et \Bz$ which must be zero when
$d_\et\Bz=0$. Hence $\Bk$ must be so chosen that
\be
d(e^{2k}h^2)/d\et = e^{2\Bk}d/d\et[F\BR'^2],
\lb{55}
\ee
when
\be e^{2k}h^2=e^{2\Bk}F\BR'^2.
\lb{56}
\ee
Dividing (\ref{55}) by (\ref{56}) and  letting $\et_t$, pronounced eta-top, be the solution for $\et$ of
\be
d_\et k =d_\et [\ln(F^{1/2}\BR'/h)]=d_\et\ln\fr{\BR'}{\BR},
\lb{57}
\ee
then evaluating (\ref{56}) at $\ze=Z,\et=\et_t$,
\be
\Bk =[ k+\ln (h/\BR')- \ha \ln F ]_{Z,\et_t}=\{k-\ln[(\BR'/\BR)\sinh(Z)]\}_{Z,\et_t}.
\lb{58}
\ee
Thus we evaluate $\Bk$ and ensure that the two gamma metrics fit on the torus.
As stated in the introduction equations (\ref{14}), (\ref{15}) and (\ref{16}) constitute the complete
 set of Einstein's equations for stationary space-times for which the Killing vector is time-like. As those equations hold both inside and outside matter and do not mention the vector potential $\bold{A}$ as opposed to the gravomagnetic field $\bold{B}$, the boundary conditions implied by them for shell distributions of matter have that property too. The general fitting procedure of Israel 
involves the vector potential so it can be simplified for stationary metrics. 
These boundary conditions arise from integrating the Einstein equations (\ref{14}), (\ref{15}) and (\ref{16}) across the bounding torus $\ze=Z$.
We integrate (\ref{14}) and use the continuity of $\xi$ the coefficient of $dt^2$ to find  the jump in the gradient of $\psi$ along the normal, denoting the integrated $T_{00}$ by $\tau_{00}$ etc. Although there is a step in the value of $\BB^2$ it does not itself have a delta-function so it does not  contribute to the integral across the surface. 
\be
-e^{-2\psi}\int{\bf \na}\psi.d{\bf S}=\int R_{00} \sqrt{-g}d^3x=\ka(\tau_{00}-\ha \xi^2 \tau)dS
\lb{59}
\ee
so
\be
-e^{-2\psi}[{\bf n.\na} \psi] =\ka [\tau_{00}-\ha \xi^2 \tau]=\ka \si_{00}.
\lb{510}
\ee
This is the generalization of the electrical $[{\bf n.E}] = 4 \pi \si.$ Evaluating the \lhs
\be
e^{-3\psi-k}h^{-1}\di\psi/\di\ze+\2e^{-2\psi}\B{n}_1d\ln{F}/d\BR=e^{-k}h^{-1}[F^{3/2} \sinh{Z}\psi_u-4Ie^{\Bk}\B{z}'/\BR]=\ka \si_{00}
\lb{511}
\ee
 where $\B{n}_1=-\B{z}'/\B{n}$ and $ \B{n}^2=(e^{-\B{k}}\B{z}')^2+F \B{R}'^2=(e^{k-\B{k}}h)^2$.
 The integral form of (\ref{15}) is $\int{\bf \bHH\!\!\cd \!dl}=-2\ka\int{\bf J\!\cd \! dS}$, where ${\bf dl}$ lies along the boundary of any chosen surface S. The only non-zero component is found by applying this to an elemental thin surface that cuts 
 a small piece of the torus's surface orthogonally at constant $\et$, there is only a contribution to the line integral from inside because $\HH$ is zero outside. So, remembering that $x^3=\eta$,
 \be
 \xi^3\BB_\phi=\HH_\phi=8Ie^{-\Bk}=2\ka \xi \tau_0^\et \sqrt{\ga_{\et \et} }\BR.
 \lb{512}
 \ee
 This is the exact analogue of the relationship between the discontinuity of the surface component of magnetic field and the surface current in electrodynamics.
 The line pressure in the $\vf$ direction is $\ka p_{\vf}=\ga_{22}T^{22}$ which can be evaluated by integrating (\ref{16}) through the surface of the torus. A similar calculation gives the sum $\ga_{kl}T^{kl}$, but to evaluate both of these we need expressions for the integrals of the components of $P^k_l$, the spatial curvature, that are transverse to the normal to the torus. Using Israel's method applied in the gamma 3-space these can be found from the external curvatures of the torus in the external and internal spaces between which it lies $\ka^a_b,~\B{\ka}^a_b$. Following Israel for  $a,b$ transverse to the normals, (we take the normals to point into the volumes in which the external curvature is calculated, hence the result is a sum rather than a difference of ${\ka}$'s).
 \be
\int P^a_b\sqrt{\ga_{\ze \ze}}d\ze=\ka^a_b+\B{\ka}^a_b.
\lb{513}
\ee
  In integrating  (\ref{16}) across the torus, the $\xi^{;k}$ only has discontinuities along the normal since the potential is continuous, so $\xi^{;k;l} $ does not contribute to the purely transverse components of the delta function on the \rhs$\!\!$.  The contribution of the first term may be evaluated following Israel's method but one dimension lower since $P^{kl}$ is the Ricci tensor of the 3-metric of $\ga$-space.
To apply Israel's formalism we calculate the external curvatures of the torus in the two gamma-spaces, with the normals pointing into each space in turn, and then add the results. We take 
$\theta^a=(\vf,\eta), a=2,3$ to be the coordinates on the torus itself. 

Outside:

The normal in the external space is along $-\bna\ze$ so $ n_k=(-e^{\psi+k}h,0,0)$ in the gamma-metric $\ga_{kl}dx^kdx^l=e^{2\psi}(e^{2k}h^2d\zeta^2+R^2d\phi^2+e^{2k}h^2d\eta^2)$. The external curvature is
\be
\ka_{ab}=- \fr{\di x^k}{\di\te^a}\fr{\di x^i}{\di\te^b}n_{k;l}=-\fr{\di x^k}{\di\te^a}\di_bn_k + \fr{\di x^k}{\di\te^a}\fr{\di x^i}{\di\te^b}\la^m_{kl}n_m,
\lb{514}
\ee
where the $\la^m_{kl}$ are the affine connections of the gamma-space. Since the normal has only a first component the first of the two terms on the \rhs vanishes and we need only calculate the $\la^1_{kl}$ with $k,l=2,3$. The only surviving terms of this type are,
\be
\la^1_{22}=-\ha e^{-2\psi-2k}h^{-2}\di_\ze(e^{2\psi}R^2)~~~;~~~\la^1_{33}= - \di_\ze(\psi+k+\ln{h})/\di{\ze}. 
\lb{515}
\ee   
From these we deduce $\ka_2^2=\la_{22}^1n_1e^{-2\ps}/R^2$ and $\ka_3^3=\la_{33}^1n_1e^{-2\ps-2k}/h^2$
\be
\ka_2^2=e^{-\psi-k} \sqrt{u_s^2-1}[\psi_u(u_s-v)+\frac{1-u_sv}{(u_s^2-1)}];
\lb{516}
\ee
\be
\ka_3^3=e^{-\psi-k}\sqrt{u_s^2-1}[(\psi_u+k_u)(u_s-v)-1].~~~
\lb{517}
\ee

Inside: 

Within the torus the gamma-metric is $\B{\gamma}_{kl}d\Bx^kd\Bx^l=e^{2\Bk} d\BR^2+\BR^2d\vf^2+F^{-1}d\Bz^2$,   where $\Bx^k=(\BR,~\vf,~\Bz)$ and on the torus it is $\BR^2d\vf^2+Hd\et^2$, where $H=e^{2\Bk} \BR'^2+\Bz'^2/F$. The normal that points into the torus is $\B{n}_k=(\B{n}_1,~0,~\B{n}_3)=(-\Bz',~0,~\BR')/\B{n}$. Where $\B{n}$ is the quantity defined under (\ref{511 st}) and
      
\be
\B{\ka}_{ab}= -\fr{\di \Bx^k}{\di\te^a}\di_b{\B n}_k + \fr{\di \Bx^k}{\di\te^a}\fr{\di \Bx^i}{\di\te^b}
\B{\la}^m _{kl} \B{n}_m.
\lb{518}
\ee
As $\B{n}_k$ has components in both the 1 and 3 directions, we now need both $\B{\la}^1_{kl}$ and $\B{\la}^3_{kl}$ with $k,l=2,3$. However $\B{\ga}_{11}$ is now constant and both $\B{\ga}_{22}$ and $\B{\ga}_{33}$ depend only on $\BR$ therefore $0=\B{\la}^1_{1l}=\B{\la}^1_{23}=\B{\la}^3_{33}=\B{\la}^3_{22}=\B{\la}^3_{11}$. We need
\be
\B{\la}^1_{22}=-e^{-2\Bk}\BR; ~~~~~\B{\la}^1_{33}=\ha e^{-2\Bk}\fr{1}{F^2}\fr{dF}{d\BR};~~~~~\B{\la}^3_{13}=\B{\la}^3_{31}=-\ha \fr{d\ln F}{d\BR}
\lb{519}
\ee
Thus we find~~~~~~~~~ $\B{\ka}_{22}=\B{\la}_{22}^1\B{n}_1;~~~~~~~\B{\ka}_{33}=-(\BR'\B{n}'_1+\Bz'\B
{n}'_3)+2\BR'\Bz'\B{\la}^3_{13}\B{n}_3+
\Bz'^2\B{\la}^1_{33}\B{n}_1.~~~~~~~$ Now $-(\BR'\B{n}'_1+\Bz'\B
{n}'_3)=(\B{z}'/\B{R}')' (\B{R}')^2/\B{n}$ so evaluating
$\B{\ka}_2^2$ and $\B{\ka}_3^3$ we obtain,
\be
\B{\ka}_2^2=\fr{e^{-2\Bk} \Bz'}{\BR \B n}:~~
\B{\ka}_3^3=e^{-2\ps-2k}h^{-2}\LLL       \lll\fr{\B{z}'}{\B{R}'}\rrr' (\B{R}')^2+2\BR'^2\Bz'\B{\la}^3_{13}-
\Bz'^3\B{\la}^1_{33}\RRR/\B{n}.
\lb{520}
\ee
The integral across the surface of the sum of the transverse spatial curvatures is found from (\ref{16}),
\be
\ka^2_2+\B{\ka}^2_2+\ka^3_3+\B{\ka}^3_3=\ka (\ga_{kl} \tau^{kl}+ \tau)=\fr{\tau_{00}}{\xi^2}.
\lb{521}
\ee
So $\tau_{00}$ is determined and hence $\tau$ is known from (\ref{59}).
Using the equation above, the integral of the 22 component of (\ref{16}) yields on multiplication by $\ga_{22}$
\be
\ka p_{\vf}= \ka_2^2+\B{\ka}^2_2-\2 \ka\tau
\lb{522}
\ee
The surface energy density, $\si$ and the $p_{\et}$ principal components of the surface stress, are related to the components of the surface energy tensor through the relationship
\be
\tau=\si - p_{\vf} - p_{\et}
\lb{523}
\ee
and those that depend on the velocity $v$ which is in the $\et$ direction,
\be
\fr{\tau_{00}}{\xi^2} = \fr{\si+p_{\et}}{1-v^2} - p_{\et}~~~;~~~\fr{\tau_0^3}{\xi }=  \fr{(\si+p_{\et})v}{1-v^2}.
\lb{524}
\ee
With $p_{\vf}$ known, these three may be solved for $v,p_{\et},$ and $\si$ successively giving,
\be
v=\fr{W}{1+\sq{1-W^2}}~~~; ~~~W=\fr{\tau_0^3/\xi}{\tau_{00}/\xi^2 -\2 (\tau+p_{\vf})},
\lb{525}
\ee
and
\be
p_{\et}=\fr{(1-v^2)(\tau_{00}/\xi^2)-(\tau+p_{\vf})}{1+v^2}~~~;~~~ \si=\tau+p_{\vf}+p_{\et}.
\lb{526}
\ee
The dominant energy conditions are $\si\ge |p_{\vf}|$ and $\si\ge|p_{\et}|$.

\section{Exploration of the Rolling Tori }
 \setcounter{equation}{0}
\subsection{ A relativistic example}
Apart from an overall scaling there is a three dimensional set of solutions. We take the 
radius of the line torus as our unit of length, $GM/c^2$ gives another length while the total mass flux $I$ is a mass per unit time  which becomes dimensionless when we use $ct $ for the time.
We find it convenient to use not the total mass $M$ itself but our quantity $m$ that is more closely 
related to the potential inside the torus. In practice $M$ is within 10\% of $m$.   There will be limits on the size of both the mass and the mass flux but these we shall explore. Since the radius of the 
line torus is our unit of length, the mass will be limited by the condition that it is supported by pressure and does not collapse into a black hole. Likewise a large mass flux can only be
held in by a torus of small minor diameter. Given the two  parameters $m$ and $ I $ we may still choose the minor axis on the torus on whose surface the mass moves. It can be small giving us a narrow tube or large like a bulky cored apple. Again there are limits on the minor axis caused by the energy conditions on the principal stresses and  the surface density which must be positive.
These imply that the velocity $v/c$ does not exceed unity.
\begin{figure}[ht].

   \includegraphics[width=7 cm]{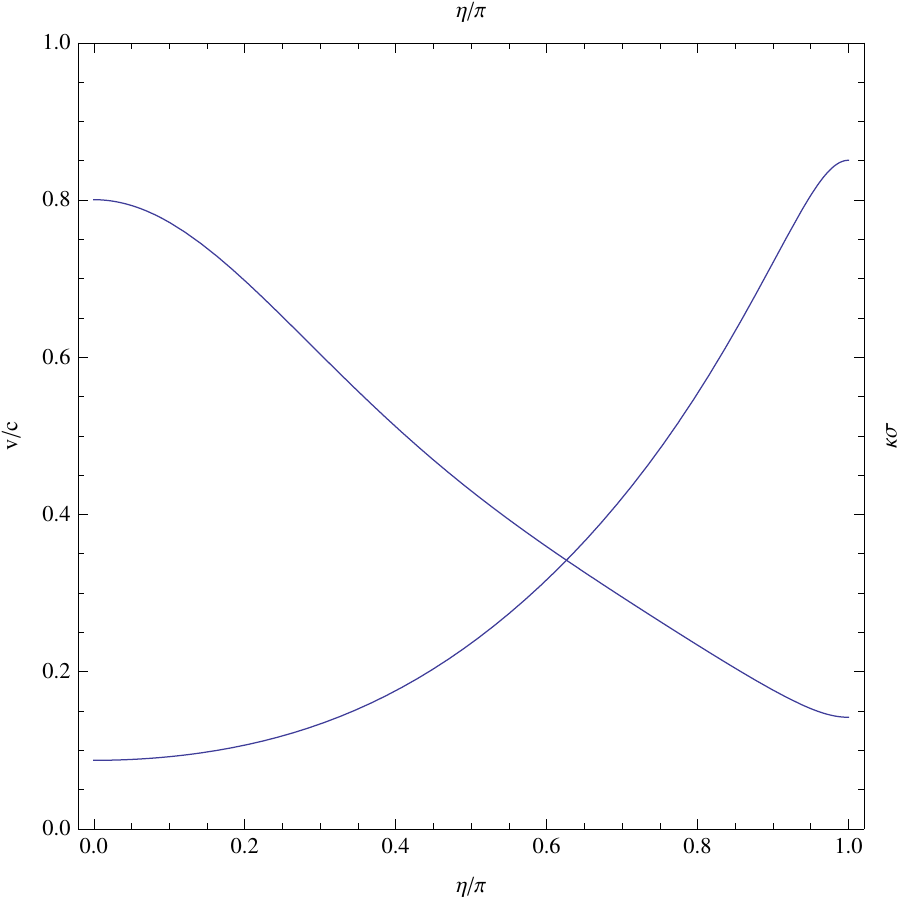} 
    
  {\small  Figure 6: Properties of the torus with $m/a=0.659, I=0.008, b/a=0.297$. The rolling velocity rises to $0.85 c$ as $\et$ increases from $0$ to $\pi$, but the positive surface density decreases over that range. This system is strongly relativistic.}
   \end{figure}
\begin{figure}[ht].

   \includegraphics[width=7 cm]{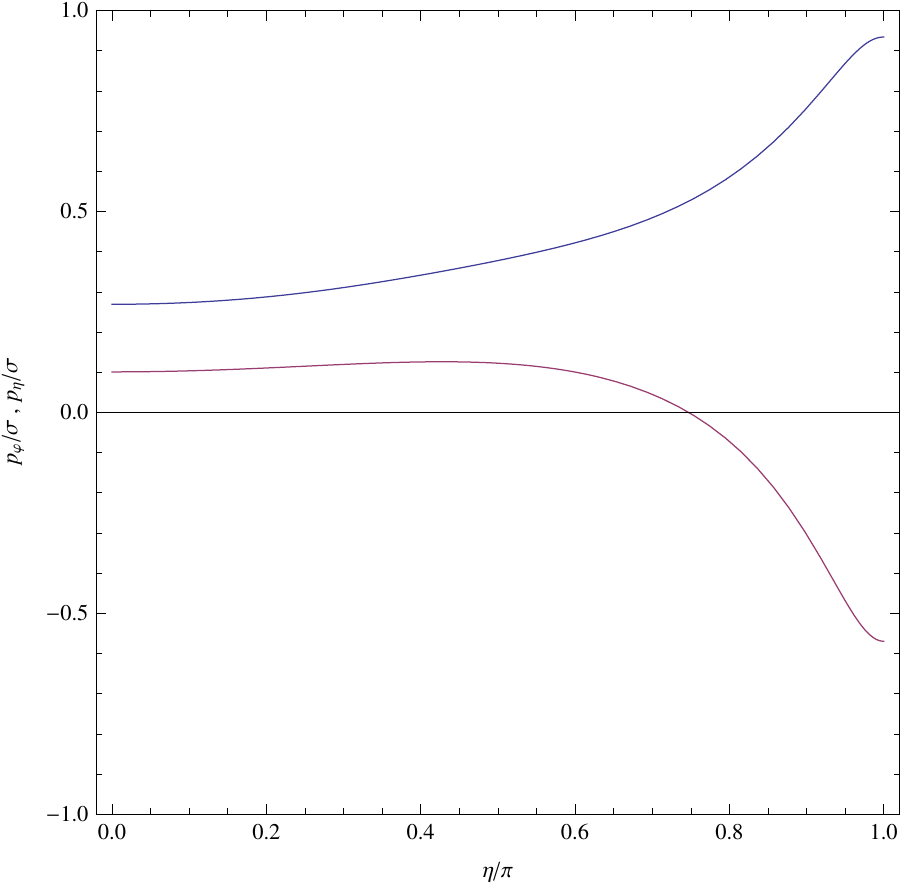} 
    
  {\small  Figure 7: The energy conditions for the same torus as Figure 6. The ratios of the principal pressures to the surface density as functions of $\et$ must remain between $\pm1$. These are satisfied but only just as the maximum is $0.934$. Notice that $p_\et$ becomes negative, (a tension) helping gravity to balance the high centrifugal force where the rolling velocity is large. }
   \end{figure}

	Figures 6 and 7 describe a strongly relativistic torus with $m/a=0.70,~~ I = 0.008$ and a minor semi-axis of $(1-u_s^2)^{-1/2}a=0.297$. The $\ze$ on this torus is $1.93$. Figure 6 gives the velocity $v$ in the $\et$ direction as a function of $\et$. It gets to $0.85$ at the inner equator but it must be emphasized that this velocity is not along the equator but at right angles to it in the $\et$ direction. This maximum  occurs on the inner equator but this coincides with the minimum surface density, $\si$, that is also portrayed in figure 6. It is  positive as it should be. Figure 7(a) gives the ratios  $p_{\vf}/\si,~~p_{\et}/\si$ which must lie between one and minus one if the dominant energy condition is to be satisfied. $p_{\vf}/\si $ reaches $0.934 $ at the inner equator so it is only just below the limit showing that we have a highly relativistic system. Notice that $p_{\et}$ starts positive but declines through zero at $\et=3\pi/4$ whereafter it becomes significantly negative showing that tension  is necessary to supplement gravity in opposing the centrifugal force due to the fast rolling motion near the inner equator. We now give the other parameters of this particular system.
Outer and inner equatorial radii $a_o/a= 1.21;~~a_i/a=0.826.$ Fourier coefficients 
$a_i =\{
0.445,~~0.0205,~~4.44\ti 10^{-4}	, ~~ 9.32 \ti10^{-6}  \}$, $i=0, 3$,
the true total mass    $M/a=0.659$ as compared with $m/a=0.7$. Notice that this is a higher total mass than the maximum that can be supported as a Bach-Weyl toroid, or as a static torus (see figure 8). The conicity  is $\B{k}=+0.0192$ and  $\B{z}$ achieves its maximum at $\et=0.361\pi$.

\subsection{Tests of the computation in the classical limits}	
	Of course our solutions which are worked out via the relativistic computation of $k$ also contain classical cases when the potential is everywhere small and the velocities much less than $1$. These provide useful checks on our numerics because in the static case the pressures times the external curvatures in their directions must balance the gravitational pull on the density. This balance gives
\be
{\rm test:}~~~(\ka_2^2  p_{\vf}/\si+\ka_3^3 p_{\et}/\si)/(\ka\si/4)=1
\lb{61}
\ee
 This test is satisfied to an accuracy of better than  half a percent for all $\et$ on a static torus with $\tm=0.001$. A similar test  but with a velocity is also satisfied when  the static test  is strongly violated for this classical system whose velocities exceed $\sq{p/\si}$. Here we must use our dynamic vtest which is,
\be
{\rm vtest:}~~~ (\ka_2^2  p_{\vf}/\si+\ka_3^3(p_{\et}/\si+v^2))/(\ka\si/4)=1
\lb{62} 
\ee
This test on a system with $\tm=0.001$ and $I=0.0001$ with a maximum velocity of $v/c= 0.025$ is 
satisfied to a similar accuracy even when the static test (applied wrongly to this dynamic case) is strongly violated. Evidently inclusion of the centrifugal force makes a very considerable difference.

\begin{figure}[ht].

   \includegraphics[width=10 cm]{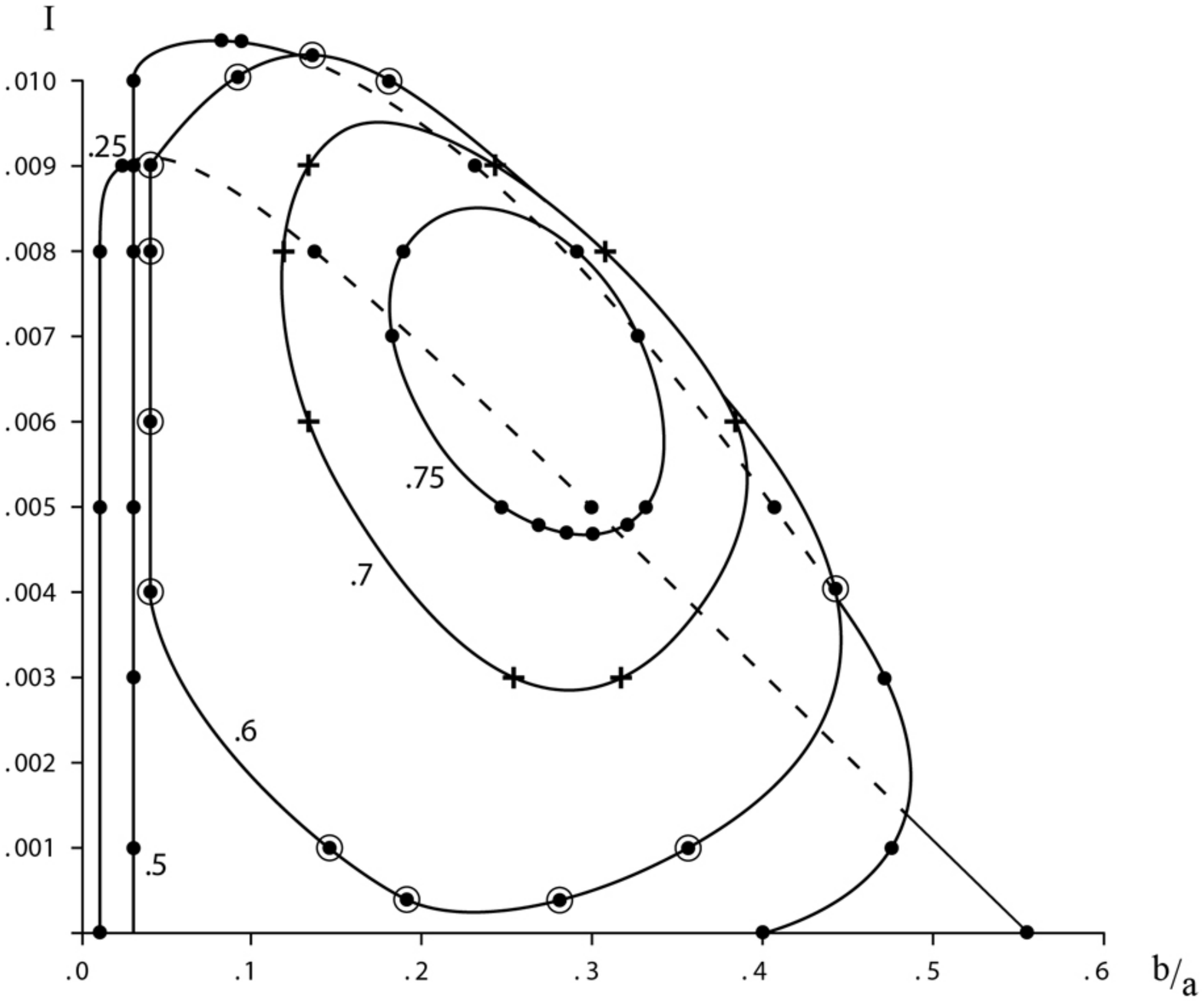} 
    
  {\small  Figure 8: Contours at $m/a = 0.25, 0.5, 0.6, 0.7, 0.75$ of the limiting surface in the $I, b/a, m/a$ space. Within the contours there are solutions obeying the dominant energy condition. Notice that the higher contours overhang the lower ones at the upper right where the lower ones are dashed. In most places the limit comes from $p_\vf/\si=1$ except at the upper left where it comes from $p_\et/\si= - 1$. Symbols denote computed points.}
   \end{figure}

\subsection{Relativistic limits due to the energy conditions}

Figure 8 shows the region in $b/a,I,m/a$-space within which the solutions satisfy the dominant energy conditions. The contours are at constant $m/a$. Evidently the rolling motion allows greater values of $m/a$ than those available for static toroids. Notice that the contours at larger $m/a$ overhang at the upper right those at lower $m/a$ so the contours actually cross. We have not seen a topographical map where this happens but wherever a mountain has been undercut by a glacier leaving an overhang it should be expected. Where the lower contours are overlaid by higher ones
they are dotted. Over most of the diagr  am it is the $p_{\vf}/\si=1$ that determines the limiting surface but this is replaced by $p_{\et}/\si=-1$ at small $b/a$ and large $I$. In all cases the surface density falls as  $\et$ increases  so the highest surface densities are achieved on the outside
equator. Generally the velocity increases inwards but for thin tori this can go the other way and
there are even cases in which the maximum in the velocity is not on the  equatorial plane. When
the velocities are not very high both principal pressures are p sttositive but at high rolling velocities
$p_{\et}$ becomes negative, so tension is required as well as gravity to hold the matter in against
the centrifugal force of the rolling motion. 
	We decided not to burden the reader with a full panoply of different cases but to give  figure 8
whose determination required many solutions to be run in the neighbourhood of this bounding surface. At the bottom of figure 8 the static tori at fixed $m/a$ do not extend to $b/a=0$. This again
demonstrates that static line tori disobey the energy conditions. Looking a bit higher up we see this is also the case for our rolling tori, in both cases $p_{\vf}/\si$ becomes too large.
	In choosing Bonnor's metric inside our torus we imposed a structure on the internal potential
that is not that which would be found if the torus were cut and unrolled into a cylinder. Thus there
will be many more solutions for rolling tori that are not included in the particular set we have chose stn. However the simplicity of Bonnor's solution suggests that those investigated here will be among the simplest solutions of this type.
\vs
   {\bf  Acknowledgments}
   \vs
 We thank Ji\v r\'\i ~Bi\v c\'ak for his interest and advice and John Harper for a reference. We particularly admired the meticulous referee who rightly rejected our earlier paper on the statics only. Nevertheless the Mathematica computations there (reproduced here) were correct. We aspire to meet his high standard of accuracy and typography in this more complicated problem.


\vskip .5 cm
\cl{\LARGE{{\bf Appendix}}}
\begin{appendix}
\setcounter{equation}{0}
\section{  Evaluation of the indefinite integrals involved in \eee (\ref{311}) }
\renewcommand{\theequation}{\Alph{section}.\arabic{equation}}
\normalsize
\setlength{\baselineskip}{20pt plus2pt}
Our general solution for $k$ was given in the unwieldy form (\ref{311}) in terms of indefinite integrals of polynomials times products of Legendre functions. Here we show how it may be expressed in terms of Legendre functions themselves. The recurrence relation $(u^2-1)P_L'=L(uP_L-P_{L-1})$ may be used to eliminate derivatives with respect to $u$ in the final results.

We first show that all eight  integrals may be reduced to known functions together with $\int P_LP_Mdu$ and $\int P_LP_Mudu$ and then evaluate those integrals. The second, fifth, seventh and eighth terms   on the right of (\ref{311}) are already of the above form. The third is, integrated by parts,
\be
\int (P_LP_M)'(u^2-1)du=P_LP_M(u^2 - 1) - 2\int P_LP_Mudu
\lb{701}
\ee
which is of the desired form. The first term   gives two integrals both of which are $l,m$ symmetric. Integrating by parts and using (\ref{701})  and Legendre's \eee$\!$ (\ref{211}),
\ba
&&\int P_L'P_M'(u^2-1)udu=P_LP_M'(u^2-1)u - (m^2-\4)\int P_LP_Mudu - \int P_LP_M'(u^2-1)du\nn
\\
&&~~~~~~=\2\!(P_LP_M)'(u^2-1)u - \2(l^2+m^2-\2)\int P_LP_Mudu -\2\int (P_LP_M)'(u^2-1)du\nn
\\
&&~~~~~~~~~~=\2(u^2-1)\LLL   (P_LP_M)'u - P_LP_M \RRR + \LLL 1-\2(l^2+m^2-\2)    \RRR\int P_LP_Mudu,\nn\\
\lb{A2}
\ea 
which is of the desired form, and
\ba
&&\int P_L'P_M'(u^2-1)du=P_LP_M'(u^2-1) - (m^2-\4)\int P_LP_Mdu\nn
\\
&&~~~~~~~~~~~=\2 (P_LP_M)'(u^2-1) - \2( l^2+m^2-\2 )\int P_LP_Mdu.
\lb{A3}
\ea
The fourth and sixth term get the required form as follows:
\be
\int P_L'P_M(u^2-1)du= P_LP_M(u^2-1) - 2\int P_LP_Mudu - \int P_LP'_M(u^2 -1) du,
\lb{A4}
\ee
which may be written
\be
2\int P_L'P_M(u^2-1)du= P_LP_M(u^2-1) - 2\int P_LP_Mudu - \int (P_LP'_M-P_L'P_M)(u^2 -1) du
\lb{A5}
\ee
The last term can be simplified because from Legendre's \eee for $P_L$ and $P_M$,
\be
\LLL  (u^2-1)(P_LP'_M - P_L'P_M)          \RRR' = -(l^2-m^2)P_LP_M.
\lb{A6}
\ee
Thus,
\be
\int(P_LP_M' -P_L'P_M)(u^2-1)du=(P_LP_M' -P_L'P_M)u(u^2-1)+(l^2-m^2)\int P_LP_Mudu.
\lb{A7}
\ee

\be
2\int P_L'P_M(u^2-1)du= (u^2-1)[  \und{P_LP_M}_{S}-  \und{(P_LP_M' - P_L'P_M)}_{A}u  ] -(\und{l^2-m^2}_{A}+\und{2}_{S})\int P_LP_Mudu.
\lb{A8}
\ee
Where symmetrical and anti-symmetrical terms for $l, m$ interchange are indicated. Taking account of symmetry and anti-symmetry in the coefficients of $m$ in the 4th and 6th terms of terms of (\ref{311}) we symmetrize   for the $l, m$ interchange and obtain their contribution to the coefficients of $a_la_m$.
To evaluate $\int P_LP_Mdu$ we integrate $P_M$ times \eee (\ref{211}) and subtract the result with $L$ and $M$ interchanged  to obtain
\be
(l^2 - m^2)\int P_LP_Mdu= - (u^2-1)(P_LP_M' - P_L'P_M)=(l-m)uP_LP_M-LP_{L-1}P_M+MP_LP_{M-1}.
\lb{A10}
\ee
This gives us the desired integral whenever
$l\ne m$. The $l=m$ integral is given by differentiating with respect to $L$ and putting $M=L$:
\be
2l\int P_L^2du=uP_L^2-P_{L-1}P_L-(l-\2)\lll P_L\fr{dP_{L-1}}{dL}-\fr{dP_L}{dL}P_{L-1}\rrr.
\lb{A11}
\ee
To evaluate $\int P_LP_Mudu$ we add $(u^2-1)P_L'P_M$ and  $(u^2-1)P_M'P_L$ , integrate 
by parts, and use the recurrence relations to obtain
\be
(l+m+1)\int P_LP_Mudu= P_LP_M(u^2-1)+ (l - \2)\int P_{L-1}P_Mdu + (m - \2)\int P_LP_{M-1}du.
 \lb{A12}
 \ee
Hence, 
\be
\int \!\!P_LP_Mudu=\fr{1}{l+m+1}  \LLL(u^2-1)   P_LP_M +  (l-\2)\int   P_{L-1}P_Mdu +    (m-\2)\int P_LP_{M-1}du\RRR. 
\lb{A13}
\ee
which are evaluated in (\ref{A10}) and (\ref{A11}).

The different terms in \eee (\ref{311}) have different dependencies on $\eta$. We therefore collect the terms in $\cos(n\eta)$ for different values of $n$ involved. 

Collecting our results  $k$ can be written in terms of $k^n_{l,m}(u)$ with $n=\pm1$ or $0$:
\ba
&k=\fr{1}{8}\sum_{l=0}^{\inf}\sum_{m=0}^{\inf}a_la_m\LLL   c(l+m+1)k^1_{l,m} +c(l+m)k^0_{l,m}+ c(l+m-1)k^{-1}_{l,m} \RRR\nn+
\\
&+\fr{1}{8}\sum_{l=0}^{\inf}\sum_{m=0}^{\inf}a_la_m\LLL   c(l-m+1)k^1_{l,-m} +c(l-m)k^0_{l,-m} +c(l-m-1)k^{-1}_{l,-m} \RRR~~
\lb{A15}
\ea
where
\ba
&&k^1_{l,m}=  -(u^2-1)\LLL    (P_LP_M)'u+(l+m) P_LP_M+(l-m)(P_LP_M'-P_L'P_M)  u    \RRR\nn
\\
&&~~~~~~~~ -\Big{\{}\LLL(l - m)^2-1\RRR(l+m-1)\Big{\}}\int _1^u P_LP_Mudu.
\lb{A16}
\ea
\be
k^0_{l,m}=2\LLL   (P_LP_M)'(u^2-1) - (l-m)^2\int _1^u P_LP_Mdu     \RRR\!.
\lb{A17}
\ee
 \ba
&&k^{-1}_{l,m}= - (u^2-1)\LLL     (P_LP_M)'u-(l+m) P_LP_M-(l-m)(P_LP_M'-P_L'P_M) u     \RRR\nn
\\
&&~~~~~~~~+\Big{\{}[(l-m)^2-1](l+m+1)\Big{\}}\int _1^uP_LP_Mudu,
\lb{A18}
\ea
The coefficients below have -m written for m in the above in  leaving L and M unchanged;
\ba
&&k^1_{l,-m}=  -(u^2-1)\LLL     (P_LP_M)'u + (l-m) P_LP_M+(l+m)(P_LP_M'-P_L'P_M) u     \RRR\nn
\\
&&~~~~~~~~~~ - \Big{\{}[(l+m)^2-1] (l-m-1)\Big{\}}\int _1^u P_LP_Mudu.
\lb{A19}
\ea
\be
k^0_{l,-m}=2\Big{\{}   (P_LP_M)'(u^2-1) - (l+m)^2 \int _1^u P_LP_Mdu    \Big{\}}\!.
\lb{A20}
\ee
Notice that when $l=m$ we need the final integral except when $l$ or $m$ is zero. 
\ba
&&k^{-1}_{l,-m}= - (u^2-1)\LLL     (P_LP_M)'u -(l-m) P_LP_M-(l+m)(P_LP_M'-P_L'P_M) u     \RRR\nn
\\
&&~~~~~~~~ +\Big{\{}[(l+m)^2-1](l-m+1)\Big{\}}\int _1^u P_LP_Mudu,
\lb{A21}
\ea
which completes our  calculation of $k^n_{l,m}$ which have been checked numerically
via numerical integration of equation (\ref{39}) at constant $\et$ and Fourier transformation.
For the rolling tori the $a_i$ fall rapidly as $i$ increases so only a few terms are needed for
accuracies of $1$ part in $10,000$.

\end{appendix}

\end{document}